\documentclass[a4,11pt]{article}
\usepackage{graphicx} 
\usepackage{amsmath}
\usepackage{amsfonts}
\usepackage{latexsym}
\usepackage{amssymb}
\makeatletter
 
  \@addtoreset{equation}{section}
 \makeatother

\begin{document}
\title{Perturbative Expansion of FBSDE in an Incomplete Market with 
Stochastic Volatility~\footnote{
This research is supported by CARF (Center for Advanced Research in Finance) and 
the global COE program ``The research and training center for new development in mathematics.''
All the contents expressed in this research are solely those of the authors and do not represent any views or 
opinions of any institutions. 
The authors are not responsible or liable in any manner for any losses and/or damages caused by the use of any contents in this research.
}}
\vspace{30mm}
\author{Masaaki Fujii,~~
Akihiko Takahashi
\\\\\\\\
Graduate School of Economics \\
The University of Tokyo\\
7-3-1 Hongo, Bunkyo-ku \\
Tokyo, Japan, 113-0033
}
\date{
First version: February 3, 2012\\
~This version: June 25, 2012
}
\maketitle



\newtheorem{definition}{Definition}
\newtheorem{assumption}{$[$ A}
\newtheorem{condition}{$[$ C}
\newtheorem{lemma}{Lemma}
\newtheorem{proposition}{Proposition}
\newtheorem{theorem}{Theorem}
\newtheorem{remark}{Remark}
\newtheorem{example}{Example}
\newtheorem{corollary}{Corollary}
\def\n{{\bf n}}
\def\A{{\bf A}}
\def\B{{\bf B}}
\def\C{{\bf C}}
\def\D{{\bf D}}
\def\E{{\bf E}}
\def\F{{\bf F}}
\def\G{{\bf G}}
\def\H{{\bf H}}
\def\I{{\bf I}}
\def\J{{\bf J}}
\def\K{{\bf K}}
\def\L{{\bf L}}
\def\M{{\bf M}}
\def\N{{\bf N}}
\def\O{{\bf O}}
\def\P{{\bf P}}
\def\Q{{\bf Q}}
\def\R{{\bf R}}
\def\S{{\bf S}}
\def\T{{\bf T}}
\def\U{{\bf U}}
\def\V{{\bf V}}
\def\W{{\bf W}}
\def\X{{\bf X}}
\def\Y{{\bf Y}}
\def\Z{{\bf Z}}
\def\cala{{\cal A}}
\def\calb{{\cal B}}
\def\calc{{\cal C}}
\def\cald{{\cal D}}
\def\cale{{\cal E}}
\def\calf{{\cal F}}
\def\calg{{\cal G}}
\def\calh{{\cal H}}
\def\cali{{\cal I}}
\def\calj{{\cal J}}
\def\calk{{\cal K}}
\def\call{{\cal L}}
\def\calm{{\cal M}}
\def\caln{{\cal N}}
\def\calo{{\cal O}}
\def\calp{{\cal P}}
\def\calq{{\cal Q}}
\def\calr{{\cal R}}
\def\cals{{\cal S}}
\def\calt{{\cal T}}
\def\calu{{\cal U}}
\def\calv{{\cal V}}
\def\calw{{\cal W}}
\def\calx{{\cal X}}
\def\caly{{\cal Y}}
\def\calz{{\cal Z}}
%
\def\sskip{\hspace{0.5cm}}
\def\simleq{ \raisebox{-.7ex}{\em $\stackrel{{\textstyle <}}{\sim}$} }
\def\leqsim{ \raisebox{-.7ex}{\em $\stackrel{{\textstyle <}}{\sim}$} }
\def\ep{\epsilon}
\def\half{\frac{1}{2}}
\def\iku{\rightarrow}
\def\Iku{\Rightarrow}
\def\ikup{\rightarrow^{p}}
\def\inclusion{\hookrightarrow}
\def\cadlag{c\`adl\`ag\ }
\def\up{\uparrow}
\def\down{\downarrow}
\def\doti{\Leftrightarrow}
\def\douti{\Leftrightarrow}
\def\dochi{\Leftrightarrow}
\def\douchi{\Leftrightarrow}%
\def\yy{\\ && \nonumber \\}
\def\y{\vspace*{3mm}\\}
\def\nn{\nonumber}
\def\be{\begin{equation}}
\def\ee{\end{equation}}
\def\bea{\begin{eqnarray}}
\def\eea{\end{eqnarray}}
\def\beas{\begin{eqnarray*}}
\def\eeas{\end{eqnarray*}}
%
\def\hd{\hat{D}}
\def\hv{\hat{V}}
\def\hsd{{\hat{d}}}
\def\hx{\hat{X}}
\def\hsx{\hat{x}}
\def\bsx{\bar{x}}
\def\bsd{{\bar{d}}}
\def\bx{\bar{X}}
\def\ba{\bar{A}}
\def\bb{\bar{B}}
\def\bc{\bar{C}}
\def\bv{\bar{V}}
\def\balpha{\bar{\alpha}}
\def\bbalpha{\bar{\bar{\alpha}}}
\def\combi{\l(\begin{array}{c}\alpha\\ \beta \end{array}\r)}
\def\f{^{(1)}}
\def\s{^{(2)}}
\def\ss{^{(2)*}}
\def\l{\left}
\def\r{\right}
\def\a{\alpha}
\def\b{\beta}
\def\L{\Lambda}

\def\E{{\bf E}}
\def\P{{\bf P}}
\def\Q{{\bf Q}}
\def\R{{\bf R}}

\def\calf{{\cal F}}
\def\calp{{\cal P}}
\def\calq{{\cal Q}}

\def\ep{\epsilon}
\def\part{\partial}
\def\del{\delta}

\def\yy{\\ && \nonumber \\}
\def\y{\vspace*{3mm}\\}
\def\nn{\nonumber}
\def\be{\begin{equation}}
\def\ee{\end{equation}}
\def\bea{\begin{eqnarray}}
\def\eea{\end{eqnarray}}
\def\beas{\begin{eqnarray*}}
\def\eeas{\end{eqnarray*}}
\def\l{\left}
\def\r{\right}

\newpage
\vspace{20mm}
\begin{abstract}
\vspace{20mm}
\normalsize
In this work, we apply our newly proposed perturbative expansion technique
to a quadratic growth FBSDE appearing in an incomplete market with stochastic volatility
that is not perfectly hedgeable.
By combining standard asymptotic expansion technique for the underlying volatility process,
we derive explicit expression for the solution of the FBSDE up to the third order of 
volatility-of-volatility for its level,  and the fourth order for its diffusion part
that can be directly translated into the optimal investment strategy. 
We compare our approximation with the exact solution,
which is known to be derived by the Cole-Hopf transformation in this popular setup.
The result is very encouraging and shows good accuracy of the approximation up to quite 
long maturities. Since our new methodology can be extended straightforwardly to multi-dimensional
setups, we expect it will open real possibilities to obtain explicit optimal portfolios or hedging strategies
under realistic assumptions.
\end{abstract}
\vspace{10mm}
\newpage

\normalsize
\section{Introduction}
In the last couple of decades, forward-backward stochastic differential equations (FBSDE) have attracted
significant academic interests. They were first introduced by Bismut (1973)~\cite{Bismut}, 
and then later extended by Pardoux and Peng (1990)~\cite{P-Peng} for general non-linear cases.
They were found particularly relevant for optimal portfolio and indifference pricing issues
in incomplete and/or constrained markets. Their financial applications are discussed in 
details in, for example, El~Karoui, Peng and Quenez (1997)~\cite{ElKaroui},
Ma and Yong (2000)~\cite{Ma} and a recent book edited by Carmona (2009)~\cite{Carmona}~.
Various topics regarding recursive utilities are thoroughly reviewed in the 
article written by Skiadas (2008)~\cite{Skiadas_review} and references therein.

FBSDEs have become also relevant in practical problems, too.
Intensive research on counterparty credit risk, collateral cost, funding rate asymmetry 
has made clear that one has to handle complicated FBSDEs for these problems (See, for example, 
\cite{Duffie, asymmetric_collateral, crepey}.).
Furthermore, forthcoming regulations on the balance sheets of financial firms and increasing demand of
cash collateral both for centrally-cleared and OTC trades are expected to constrain trader's position 
severely, and may even turn a part of financial products effectively nontradable.
These new developments in the financial market will make deeper understanding of FBSDEs a more 
pressing issue in the coming years.

In the previous work~\cite{analytic_FBSDE}, we have presented 
a simple analytical approximation scheme for generic non-linear FBSDEs.
By treating the interested system as the linear decoupled FBSDE 
perturbed by a non-linear driver and feedback terms, the problem of each 
order of approximation turns out to be equivalent to those for pricing of 
standard European contingent claims.
In this work, we consider its application to a particular type of FBSDEs with a quadratic growth driver.
This type of system is receiving strong attention because it
appears in the optimal portfolio problems for very popular utilities of exponential 
and power forms.
In particular, we study the optimal portfolio problem in an incomplete market 
with one risky asset whose stochastic volatility is not perfectly hedgeable.
We derive the explicit solution of the corresponding 
FBSDE up to the third order of volatility of volatility (vol-of-vol) for the first "level" component,
and the fourth order of vol-of-vol for the second "diffusion" component.
It allows us to have the explicit expression of the optimal strategy, which is 
of great importance for practical applications.

In the particular setup we use in this paper, a special transformation 
of variable known as {\it the Cole-Hopf transformation} gives the closed form
expression~\cite{Zaripho}\footnote{It still requires numerical simulation to evaluate the expectation.},
which allows us to test accuracy of the perturbative expansion for both of the 
backward components. We shall see that the comparisons to the solution are quite encouraging. 
Since our approximation scheme is easily extended 
to multi-dimensional setups, we expect it will open 
real possibilities to obtain explicit optimal portfolios or hedging strategies
in more realistic situations, which is so far limited to very simplistic models.

\section{Setup}
We consider a probability space $(\Omega, \calf, \mathbb{P})$, where
$\calf$ is the augmented filtration generated by two dimensional
Brownian motion $(B_1, B_2)$.
The market consists of one risk-free money market account with 
zero interest rate, and one risky asset with 
stochastic volatility.
The SDEs of the risky asset $S$ and its volatility $X$ are assumed to follow
\bea
dS_t/S_t&=&\mu dt+\sqrt{X_t}\Bigl(\rho dB_{1t}+\sqrt{1-\rho^2}dB_{2t}\Bigr)\\
dX_t&=&k\bigl(m-X_t)dt+c\sqrt{X_t}dB_{1t}
\eea
where $\rho \in(-1,1)$ is a constant correlation parameter and $\mu, k, m$ and $c$
are all positive constants.
Let us denote $\pi_t$ is the invested amount to the risky asset.
Then, the investor's wealth dynamics follows
\bea
dW_t^{\pi}=\mu \pi_t dt+ \pi_t \sqrt{X_t}\Bigl(\rho dB_{1t}+\sqrt{1-\rho^2} dB_{2t}\Bigr)
\eea
with the initial endowment $w_0$.
We assume that the utility of an agent is given by the exponential form with risk aversion 
parameter $\gamma>0$ and only dependent on the terminal wealth at time $T$.
Let us denote a function $U$ as
\be
U(x)=-\exp\Bigl(-\gamma x\Bigr)~,
\ee
and then the agent's problem is given by
\be
J(w_0)=\sup_{\pi\in \cala}\mathbb{E}\Bigl[U(W^\pi_T)\Bigr]
\ee
where $\cala$ is the set of all the admissible strategies.

It is well known that the above problem can be represented by a quadratic growth
FBSDE. Particularly simple and clear derivation of the relevant FBSDE
are given in Hu, Imkeller and M\"uller (2005)~\cite{Imkeller1}
for exponential and power utilities, and 
in Horst et al. (2011)~\cite{Imkeller2} for generic form of utilities.
It can be shown that the optimal strategy $\pi^{*}$ is specified by
\be
\pi_t^*=\frac{1}{\gamma X_t}\Bigl(\mu-\gamma\rho \sqrt{X_t}Z_t\Bigr)
\label{optport}
\ee
where $Z$ is a solution of the following FBSDE:
\bea
dV_t&=&-f(Z_t,X_t)dt+Z_t dB_{1t}\nn\\
V_T&=&0 
\label{FBSDEorg}
\eea
with a quadratic growth driver:
\be
f(Z_t,X_t)=-\frac{\gamma}{2}(1-\rho^2)Z_t^2-\frac{\mu}{\sqrt{X_t}}\rho Z_t+\frac{1}{2\gamma}
\frac{\mu^2}{X_t}~.
\ee
One can concentrate on the FBSDE system composed by $X$ and $V$ since the dynamics of $S$ itself
drops off from the system.
In the following, we denote $B_t$ instead of $B_{1t}$ for simplicity.

\section{Perturbative Expansion}
We now introduce a perturbative expansion parameter $\ep$
to render the original system linear decoupled FBSDE in each order of $\ep$.
We write
\bea
dV_t^{(\ep)}&=&-\frac{\mu^2}{2\gamma X_t}dt-\ep g(Z_t^{(\ep)},X_t)dt+Z_t^{(\ep)}dB_t\\
V_T^{(\ep)}&=&0
\eea
where 
\be
g(z,x)=-\frac{\gamma}{2}(1-\rho^2)z^2-\frac{\mu \rho}{\sqrt{x}}z~.
\ee
We suppose that the solution is given by a perturbative expansion in terms of $\ep$ as
\bea
V_t^{(\ep)}&=&V_t^{(0)}+\ep V_t^{(1)}+\ep^2 V_t^{(2)}+\cdots \\
Z_t^{(\ep)}&=&Z_t^{(0)}+\ep Z_t^{(1)}+\ep^2 Z_t^{(2)}+\cdots ~.
\eea
Although it is possible to eliminate the linear term of $z$ from the driver function $g(z,x)$ by 
using the change of probability measure, we treat it directly here since it is not always a practical method
in the presence of complicated state dependencies in its coefficient in more realistic 
situations.

Once we obtain the solution up to the certain order of $\ep$, then putting 
$\ep=1$ will provide a reasonable approximation as long as the contribution 
from $g(z,x)$ is small enough. In economic terms, the above approximation 
corresponds to an expansion of the optimal strategy around the 
myopic mean-variance portfolio. It is expected to be naturally fit to our perturbative assumption
as long as the hedging contribution is only sub-dominant.
In the reminder of this work, we consider the expansion up to the third order of $\ep$. 

\begin{proposition}
\label{proposition1}
$(V^{(i)},Z^{(i)})$ with $i=\{0,1,2,3\}$ follow the linear FBSDEs given below:
\bea
dV^{(0)}_t&=&-\frac{\mu^2}{2\gamma}\frac{1}{X_t}dt+Z_t^{(0)}dB_t \\
dV^{(1)}_t&=&-g(Z_t^{(0)},X_t)dt+Z_t^{(1)}dB_t\\
dV^{(2)}_t&=&-\partial_z g(Z_t^{(0)},X_t)Z_t^{(1)}dt+Z_t^{(2)}dB_t\\
dV^{(3)}_t&=&-\left\{\partial_z g(Z_t^{(0)},X_t)Z_t^{(2)}+\frac{1}{2}\partial_z^2 g(Z_t^{(0)},X_t)(Z_t^{(1)})^2
\right\}dt+Z_t^{(3)}dB_t,
\eea
where the terminal values are all zero, $V_T^{(i)}=0$ with $i\in \{0,1,2,3\}$, and 
$\part_z$ denotes partial derivative with respect to the first argument of function $g(z,x)$.
\end{proposition}
{\it Proof}: It follows from a straightforward application of the method given in \cite{analytic_FBSDE}. $\blacksquare$
\\

From Proposition~\ref{proposition1}, one can see that each pair of $(V^{(i)},Z^{(i)})$ 
is a solution of a linear decoupled FBSDE and thus easy to integrate.
One obtains\\
zeroth order:
\bea
V_t^{(0)}&=&\frac{\mu^2}{2\gamma}\int_t^T \mathbb{E}\left[\left.\frac{1}{X_u}\right|\calf_t\right]du\\
Z_t^{(0)}&=&\frac{\mu^2}{2\gamma}\int_t^T \mathbb{E}\left[\left.\cald_t\left(\frac{1}{X_u}\right)\right|\calf_t\right]du
\eea
first order:
\bea
V_t^{(1)}&=&\int_t^T \mathbb{E}\Bigl[g(Z_u^{(0)},X_u)\Bigr|\calf_t\Bigr]du\\
Z_t^{(1)}&=&\int_t^T \mathbb{E}\Bigl[\cald_t g(Z_u^{(0)},X_u)\Bigr|\calf_t\Bigr]du
\eea
second order:
\bea
V_t^{(2)}&=&\int_t^T \mathbb{E}\Bigl[\part_z g(Z_u^{(0)},X_u)Z_u^{(1)}\Bigr|\calf_t\Bigr]du \\
Z_t^{(2)}&=&\int_t^T \mathbb{E}\Bigl[\cald_t\Bigl(\part_z g(Z_u^{(0)},X_u)Z_u^{(1)}\Bigr)\Bigr|\calf_t\Bigr]du 
\eea
third order:
\bea
V_t^{(3)}&=&\int_t^T \mathbb{E}\left[\left. \partial_z g(Z_u^{(0)},X_u)Z_u^{(2)}+\frac{1}{2}\partial_z^2 g(Z_u^{(0)},X_u)(Z_u^{(1)})^2 \right|\calf_t\right]du\\
Z_t^{(3)}&=&\int_t^T \mathbb{E}\left[\left. \cald_t \Bigl(\partial_z g(Z_u^{(0)},X_u)Z_u^{(2)}+\frac{1}{2}\partial_z^2 g(Z_u^{(0)},X_u)(Z_u^{(1)})^2\Bigr)\right|\calf_t\right]du
\eea
respectively, where $\cald_t$ is a Malliavin derivative with respect to $B$. 

\section{Asymptotic Expansion}
\label{sec-expansion}
Although, in the previous section,  we have formally expanded the original non-linear FBSDE in terms of a series of 
linear decoupled FBSDEs, we need to explicitly evaluate the involved expectations to obtain a quantitative result.
As explained in \cite{analytic_FBSDE}, this can be done  by making use of standard asymptotic expansion technique, which is now widely used for pricing of various European contingent claims and 
also for computation of the optimal portfolio in complete markets 
(See, for examples \cite{KT,STT,T,asymptotic3,TY}
and references therein for concrete examples.).

We introduce a different parameter $\del$
to expand the forward component $X$ in terms of the vol-of-vol, ie, $c$:
\bea
dX_u^{(\del)}=k(m-X_u^{(\del)})du+\del c\sqrt{X_u^{(\del)}}dB_u~.
\eea
We expand $X$ up to the third order of $\del$ as 
\bea
X_u^{(\del)}&=&X_u^{(0)}+\del D_{tu}+\frac{\del^2}{2}E_{tu}+\frac{\del^3}{3!}F_{tu}+o(\del^3) \\
X_t^{(\del)}&=&x_t
\eea
where each term is defined by
\bea
D_{tu}=\left.\frac{\part X_u^{(\del)}}{\part \del}\right|_{\del=0}, \qquad
E_{tu}=\left.\frac{\part^2 X_u^{(\del)}}{\part \del^2}\right|_{\del=0}, \qquad
F_{tu}=\left.\frac{\part^3 X_u^{(\del)}}{\part \del^3}\right|_{\del=0}~.
\eea
The relevant formulas regarding the above expansions are summarized in Appendix~\ref{X_expansion}.

Now, in each order of $\ep$, we try to expand the backward components in terms of $\del$.
More concretely, we are going to approximate each pair of $(V^{(i)}, Z^{(i)})$ with $i\in\{0,1,2,3\}$ as
\bea
V_t^{(i,\del)}&=&V_t^{(i,0)}+\del V_t^{(i,1)}+\frac{\del^2}{2} V_t^{(i,2)}+\frac{\del^3}{3!}V_t^{(i,3)}+o(\del^3) \\
Z_t^{(i,\del)}&=&Z_t^{(i,0)}+\del Z_t^{(i,1)}+\frac{\del^2}{2}Z_t^{(i,2)}+\frac{\del^3}{3!}Z_t^{(i,3)}+\frac{\del^4}{4!}Z^{(i,4)}+o(\del^4)
\eea
As we shall see, the required calculation to obtain $V^{(i,j)}$ is to take expectation value of a polynomial function of $X^{(k)}$ with $k\in\{1,2,3\}$.
Since each $X^{(k)}$ is given by a multiple Wiener integral, the evaluation of the expectation for $V^{(i,j)}$
can be easily calculated.  Once $V^{(i,j)}$ is obtained explicitly in terms of $x_t$, 
simple application of It\^o's formula gives us the expression of $Z^{(i,j+1)}$ by
\bea
Z_t^{(i,j+1)}=(j+1)c\sqrt{x_t} \frac{\part}{\part x_t} V_t^{(i,j)}(x_t)~.
\eea
It is easy to see that $Z^{(i,0)}$ is zero. The reason why we expand $Z$ by one higher order 
is to study the convergence of $Z$ itself.
As long as the vol-of-vol (or $c$) is small relative to the other parameters,
putting $\del=1$ is expected to give a reasonable approximation to the original model.

\subsection{Asymptotic Expansion of $V^{(0,\del)}$}
\label{sec-V0del}
In the zero-th order of $\ep$, we want to expand
\bea
\label{V0del}
V_t^{(0,\del)}(x_t)=\frac{\mu^2}{2\gamma}\int_t^T \mathbb{E}\Bigl[v_u^{(\del)}\Bigr|\calf_t\Bigr]du
\eea
in terms of $\del$, where
\be
v_u^{(\del)}=\frac{1}{X_u^{(\del)}}~.
\ee
One can show that
\bea
v_u^{(\del)}=v_u^{(0)}+\del v_u^{(1)}+\frac{\del^2}{2}v_u^{(2)}+\frac{\del^3}{3!}v_u^{(3)}+o(\del^3)
\eea
where each term is given by
\bea
v_u^{(0)}&=&(X_u^{(0)})^{-1}\\
v_u^{(1)}&=&-(X_u^{(0)})^{-2}D_{tu}\\
v_u^{(2)}&=&2(X_u^{(0)})^{-3}D^2_{tu}-(X_u^{(0)})^{-2}E_{tu}\\
v_u^{(3)}&=&-6(X_u^{(0)})^{-4}D^3_{tu}+6(X_u^{(0)})^{-3}D_{tu}E_{tu}-(X_u^{(0)})^{-2}F_{tu}~.
\eea
Let us define
\bea
v_u^{(i)}(x_t):=\mathbb{E}\Bigl[v_u^{(i)}\Bigr|\calf_t\Bigr]
\eea
then, from the results of Appendix, one can check that
\bea
v_u^{(1)}(x_t)=v_u^{(3)}(x_t)=0
\eea
and also
\bea
v_u^{(0)}(x_t)&=&(X_u^{(0)}(x_t))^{-1}\\
v_u^{(2)}(x_t)&=&2(X_u^{(0)}(x_t))^{-3}D^2_{tu}(x_t)~.
\eea
Integration in (\ref{V0del}) can be performed explicitly as
\bea
V_t^{(0,\del)}(x_t)=V_t^{(0,0)}(x_t)+\frac{\del^2}{2}V_t^{(0,2)}(x_t)+o(\del^3)
\eea
where
\bea
V_t^{(0,0)}(x_t)&=&-\frac{\mu^2}{2\gamma}\frac{1}{k m} \ln\left(\frac{Y_{tT} x_t}{X_T^{(0)}(x_t)}\right)\\
V_t^{(0,2)}(x_t)&=&-\frac{\mu^2}{2\gamma}\frac{c^2}{k^2}\left\{
\frac{(1-Y_{tT})\bigl[m(1-Y_{tT})+2Y_{tT}x_t\bigr]}{2m (X_T^{(0)}(x_t))^2}
+\frac{1}{m^2}\ln\left(\frac{Y_{tT} x_t}{X_T^{(0)}(x_t)}\right)\right\}~.\nn \\
\eea
The relevant definitions of variables are given in Appendix.

\subsection{Asymptotic Expansion of $Z^{(0,\del)}$}
Although we have considered the dynamics of Malliavin derivative $\cald_t X_u^{(\del)}$ directly
in \cite{analytic_FBSDE}, it is easier to simply apply It\^o's formula
to the result of $V^{(0,\del)}$,
since we already have its explicit expression in terms of $x_t$. One can
easily confirm that
\bea
Z_t^{(0,\del)}(x_t)=\del Z_t^{(0,1)}(x_t)+\frac{\del^3}{3!}Z_t^{(0,3)}(x_t)+o(\del^4)
\eea
where
\bea
Z_t^{(0,1)}(x_t)&=&-\frac{\mu^2 c}{2\gamma k}\frac{1-Y_{tT}}{\sqrt{x_t}(X_T^{(0)}(x_t))}\\
Z_t^{(0,3)}(x_t)&=&-\frac{3\mu^2 c^3}{2\gamma k^2}\frac{(1-Y_{tT})^2}{\sqrt{x_t}
(X_T^{(0)}(x_t))^3}\Bigl[m(1-Y_{tT})+2Y_{tT}~x_t\Bigr]~.
\eea

\subsection{Asymptotic Expansion of $V^{(1,\del)}$}
\label{sec-V1del}
In the first order of $\ep$, we need to expand
\bea
&&V_t^{(1,\del)}(x_t)=\int_t^T \mathbb{E}\Bigl[g(Z_u^{(0,\del)},X_u^{(\del)})\Bigr|\calf_t\Bigr]du\\
&&~=-\frac{\gamma}{2}(1-\rho^2)\int_t^T
\mathbb{E}\Bigl[(Z_u^{(0,\del)})^2\Bigr|\calf_t\Bigr]du
-\mu\rho \int_t^T \mathbb{E}\Bigl[(X_u^{(\del)})^{-\frac{1}{2}}Z_u^{(0,\del)}\Bigr|\calf_t\Bigr]du~.\nn\\
\label{V1del}
\eea
From the previous results, we have
\bea
Z_u^{(0,\del)}=\del Z_u^{(0,1)}(X_u^{(\del)})+\frac{\del^3}{3!}Z_u^{(0,3)}(X_u^{(\del)})+o(\del^4)
\eea
and hence both of the integrands in (\ref{V1del}) can be explicitly written as a function of $X_u^{(\del)}$.
Therefore, we can follow the same procedures in Section~\ref{sec-V0del}:
Firstly apply $\part_\del$, ie, partial derivative with respect to $\del$, 
and then express the integrand as a function of $X_u^{(0)}$, $D_{tu}$ etc..   
The evaluation of its expectation is now easily performed using the results given in Appendix.  
After straightforward but lengthy calculation, we obtain
\be
V_t^{(1,\del)}(x_t)=\del V_t^{(1,1)}(x_t)+\frac{\del^2}{2}V_t^{(1,2)}(x_t)+\frac{\del^3}{3!}V_t^{(1,3)}(x_t)+o(\del^3)
\ee
where
\bea
V_t^{(1,1)}(x_t)&=& -\frac{\rho \mu^3 c}{2\gamma k^2}\left\{
\frac{(1-Y_{tT})}{m X_T^{(0)}(x_t)}+\frac{1}{m^2}\ln\left(\frac{Y_{tT}~x_t}{X_T^{(0)}(x_t)}\right)\right\} \\
V_t^{(1,2)}(x_t)&=&(1-\rho^2)\frac{\mu^4 c^2}{4\gamma k^3}
\left\{
\frac{(1-Y_{tT})[3m(1-Y_{tT})+2Y_{tT}~x_t]}{2m^2 (X_T^{(0)}(x_t))^2}+
\frac{1}{m^3}\ln\left(\frac{Y_{tT}~ x_t}{X_T^{(0)}(x_t)}\right)
\right\} \nn
\\\\
V_t^{(1,3)}(x_t)&=&\frac{3\rho \mu^3 c^3}{2\gamma k^3}\left\{
\frac{(1-Y_{tT})}{2m^2(X_T^{(0)}(x_t))^2}\bigl[m(1-Y_{tT})-2Y_{tT}~x_t\bigr]-\frac{2}{m^3}\ln
\left(\frac{Y_{tT}~x_t}{X_T^{(0)}(x_t)}\right)\right.\nn\\
&&\hspace{15mm}\left.-\frac{(1-Y_{tT})}{2m^2(X_T^{(0)}(x_t))^3}\Bigl[
5m^2(1-Y_{tT})^2+9m(1-Y_{tT})(Y_{tT}x_t)+2(Y_{tT}x_t)^2\Bigr]\right\}~.\nn\\
\eea

\subsection{Asymptotic Expansion of $Z^{(1,\del)}$}
By applying  It\^o's formula to the expanded $V^{(1,\del)}$, one obtains the volatility 
component easily as before:
\bea
Z_t^{(1,\del)}(x_t)=\frac{\del^2}{2}Z_t^{(1,2)}(x_t)+\frac{\del^3}{3!}Z_t^{(1,3)}(x_t)+\frac{\del^4}{4!}Z_t^{(1,4)}(x_t)+o(\del^4)
\eea
where
\bea
Z_t^{(1,2)}(x_t)&=&-\frac{\rho \mu^3 c^2}{\gamma k^2}\frac{(1-Y_{tT})^2}{\sqrt{x_t}(X_T^{(0)}(x_t))^2}\\
Z_t^{(1,3)}(x_t)&=&(1-\rho^2)\frac{3\mu^4 c^3}{4\gamma k^3}\frac{(1-Y_{tT})^3}{\sqrt{x_t}(X_T^{(0)}(x_t))^3}~ \\
Z_t^{(1,4)}(x_t)&=&-\frac{6 \rho \mu^3 c^4}{\gamma k^3}\frac{(1-Y_{tT})^3[2m(1-Y_{tT})+5 Y_{tT}~x_t]}{\sqrt{x_t}(X_T^{(0)}(x_t))^4}~.
\eea

\subsection{Asymptotic Expansion of $V^{(2,\del)}$}
In the second order of $\ep$, we have to evaluate
\bea
&&V_t^{(2,\del)}(x_t)=\int_t^T \mathbb{E}\Bigl[\part_z g(Z_u^{(0,\del)},X_u^{(\del)})Z_u^{(1,\del)}
\Bigr|\calf_t\Bigr]du\\
&&\quad=-\gamma(1-\rho^2)\int_t^T \mathbb{E}\Bigl[
Z_u^{(0,\del)}Z_u^{(1,\del)}\Bigr|\calf_t\Bigr]du-
\mu\rho\int_t^T \mathbb{E}\Bigl[(X_u^{(\del)})^{-\frac{1}{2}}Z_u^{(1,\del)}\Bigr|\calf_t\Bigr]du~.\nn\\
\eea
Following the same arguments in Section~\ref{sec-V1del}, we can express the 
above expectation explicitly. After tedious calculation, one obtains
\bea
V_t^{(2,\del)}(x_t)=\frac{\del^2}{2}V_t^{(2,2)}(x_t)+\frac{\del^3}{3!}V_t^{(2,3)}(x_t)+o(\del^3)
\eea
where
\bea
V_t^{(2,2)}(x_t)&=&-\frac{\rho^2 \mu^4 c^2}{\gamma k^3}\left\{\frac{(1-Y_{tT})[3m(1-Y_{tT})+2Y_{tT}x_t]}{2m^2 (X_T^{(0)}(x_t))^2}
+\frac{1}{m^3}\ln\left(\frac{Y_{tT}~x_t}{X_T^{(0)}(x_t)}\right)\right\}\nn
\\\\
V_t^{(2,3)}(x_t)&=&\rho(1-\rho^2)\frac{9\mu^5c^3}{4\gamma k^4}
\left\{\frac{(1-Y_{tT})\bigl[11m^2(1-Y_{tT})^2+15m(1-Y_{tT})(Y_{tT}x_t)
+6(Y_{tT}x_t)^2\bigr]}{6m^3 (X_T^{(0)}(x_t))^3}\right.\nn\\
&&\hspace{30mm}\left.+\frac{1}{m^4}\ln\left(\frac{Y_{tT}~x_t}{X_T^{(0)}(x_t)}\right)\right\}~.
\eea

\subsection{Asymptotic Expansion of $Z^{(2,\del)}$}
As before, simple application of It\^o's formula yields
\bea
Z_t^{(2,\del)}(x_t)=\frac{\del^3}{3!}Z_t^{(2,3)}(x_t)+\frac{\del^4}{4!}Z_t^{(2,4)}+o(\del^4)
\eea
where
\bea
Z_t^{(2,3)}(x_t)&=&-\frac{3\rho^2 \mu^4 c^3}{\gamma k^3}
\frac{(1-Y_{tT})^3}{\sqrt{x_t}(X_T^{(0)}(x_t))^3}~ \\
Z_t^{(2,4)}(x_t)&=&\rho (1-\rho^2)\frac{9 \mu^5 c^4}{\gamma k^4}\frac{(1-Y_{tT})^4}{\sqrt{x_t}(X_T^{(0)}(x_t))^4}~.
\eea

\subsection{Asymptotic Expansion of $(V^{(3,\del)},Z^{(3,\del)})$}
We have
\bea
V_t^{(3,\del)}(x_t)=\int_t^T \mathbb{E}\left[\left.
\part_z g(Z_u^{(0,\del)},X_u^{(\del)})Z_u^{(2,\del)}+\frac{1}{2}\part_z^2g(Z_u^{(0,\del)},X_u^{(\del)})
\bigl(Z_u^{(1,\del)}\bigr)^2\right|\calf_t\right]du
\eea
and we can easily confirm that the contribution of $O(\del^3)$ comes only from the first term.
The result is 
\bea
V_t^{(3,\del)}(x_t)=\frac{\del^3}{3!}V_t^{(3,3)}(x_t)+o(\del^3)
\eea
where
\bea
V_t^{(3,3)}(x_t)&=&-\frac{3\rho^3 \mu^5 c^3}{\gamma k^4}\left\{
\frac{(1-Y_{tT})\Bigl[11m^2(1-Y_{tT})^2+15m(1-Y_{tT})(Y_{tT}x_t)
+6(Y_{tT}x_t)^2\Bigr]}{6m^3 (X_T^{(0)}(x_t))^3}\right.\nn\\
&&\hspace{30mm}\left.+\frac{1}{m^4}\ln\left(\frac{Y_{tT}~x_t}{X_T^{(0)}(x_t)}\right)\right\}~.
\eea
It is clear to see
\bea
Z_t^{(3,\del)}(x_t)=\frac{\del^4}{4!}Z_t^{(3,4)}+o(\del^4)~
\eea
where
\be
Z_t^{(3,4)}=-\frac{12 \rho^3 \mu^5 c^4}{\gamma k^4}\frac{(1-Y_{tT})^4}{\sqrt{x_t}(X_T^{(0)}(x_t))^4}~.
\ee

\subsection{Asymptotic Expansion of $(V^{(i,\del)},Z^{(i,\del)})$ with $(i\geq 4)$}
\label{sec-V4del}
Let us consider what happens when we proceed further to a higher order of $\ep$.
In the fourth order, we see that $V^{(4,\del)}$ has contributions from
\bea
&&\part_z g(Z_u^{(0,\del)},X_u^{(\del)})Z_u^{(3,\del)}\\
&&\part_z^2 g(Z_u^{(0,\del)},X_u^{(\del)})Z_u^{(1,\del)}Z_u^{(2,\del)}\\
&&\part_z^3 g(Z_u^{(0,\del)},X_u^{(\del)})\bigl(Z_u^{(1,\del)}\bigr)^3~
\eea
where the last term vanishes and all the others have $o(\del^3)$. Therefore we have $V^{(4,\del)}=o(\del^3)$
and hence obviously, $Z^{(4,\del)}=o(\del^4)$.
By repeating the same arguments, we can conclude
\bea
&&V_t^{(i,\del)}=o(\del^3)\\
&&Z_t^{(i,\del)}=o(\del^4)
\eea
for all $i\geq 4$.

\subsection{Summary of Expansion and its Interpretation}
Let us suppose, as we have hypothesized at the beginning, that the perturbative expansions
\bea
V_t^{(\ep)}&=&V_t^{(0)}+\ep V_t^{(1)}+ \ep^2 V_t^{(2)}+ \ep^3 V_t^{(3)}+\cdots \\
Z_t^{(\ep)}&=&Z_t^{(0)}+\ep Z_t^{(1)}+ \ep^2 Z_t^{(2)}+ \ep^3 Z_t^{(3)}+\cdots
\label{pe}
\eea
really converges to the true solution.  From the previous observation, it is easy to see that there is no contribution 
to the solution of FBSDE from the fourth or higher order terms of $\ep$ as long as we work in $O(\del^3)$ for $V$
and $O(\del^4)$ for $Z$ components, respectively.
Therefore, the results we have obtained can be interpreted as the asymptotic expansion of the
true solution of the FBSDE in $O(\del^3)$ for the level component $V$ and in $O(\del^4)$ for the 
diffusion component $Z$.

As a summary, whole of the discussion in Section~\ref{sec-expansion} leads to the next
proposition:
\begin{proposition}
The solution $(V,Z)$ of the following FBSDE:
\bea
dV_t&=& -\l\{-\frac{\gamma}{2}(1-\rho^2)Z_t^2-\frac{\mu}{\sqrt{X_t}}\rho Z_t
+\frac{1}{2\gamma}\frac{\mu^2}{X_t}\r\}dt + Z_t dB_t;\ V_T=0, \\
dX_t&=&k(m-X_t)dt+c\sqrt{X_t}dB_t;\ X_0=x
\eea
can be asymptotically expanded in terms of vol-of-vol that is 
$c$, as:
\bea
V_t(x_t)&=&V_t^{(0,0)}(x_t)+\frac{1}{2}V_t^{(0,2)}(x_t)+V_t^{(1,1)}(x_t)+\frac{1}{2}V_t^{(1,2)}(x_t)+\frac{1}{3!}V_t^{(1,3)}(x_t)\nn\\
&&+\frac{1}{2}V_t^{(2,2)}(x_t)+\frac{1}{3!}V_t^{(2,3)}(x_t)+\frac{1}{3!}V_t^{(3,3)}(x_t)+o(c^3) \\
Z_t(x_t)&=&Z_t^{(0,1)}(x_t)+\frac{1}{3!}Z_t^{(0,3)}(x_t)+\frac{1}{2}Z_t^{(1,2)}(x_t)+\frac{1}{3!}Z_t^{(1,3)}(x_t)
+\frac{1}{4!}Z_t^{(1,4)}\nn \\
&&+\frac{1}{3!}Z_t^{(2,3)}(x_t)+\frac{1}{4!}Z_t^{(2,4)}+\frac{1}{4!}Z_t^{(3,4)}+o(c^4)~,
\eea
where each term is given by
\bea
V_t^{(0,0)}(x_t)&=&-\frac{\mu^2}{2\gamma}\frac{1}{k m} \ln\left(\frac{Y_{tT} x_t}{X_T^{(0)}(x_t)}\right) \nn\\
V_t^{(0,2)}(x_t)&=&-\frac{\mu^2}{2\gamma}\frac{c^2}{k^2}\left\{
\frac{(1-Y_{tT})\bigl[m(1-Y_{tT})+2Y_{tT}x_t\bigr]}{2m (X_T^{(0)}(x_t))^2}
+\frac{1}{m^2}\ln\left(\frac{Y_{tT} x_t}{X_T^{(0)}(x_t)}\right)\right\}~\nn
\eea
\bea
V_t^{(1,1)}(x_t)&=& -\frac{\rho \mu^3 c}{2\gamma k^2}\left\{
\frac{(1-Y_{tT})}{m X_T^{(0)}(x_t)}+\frac{1}{m^2}\ln\left(\frac{Y_{tT}~x_t}{X_T^{(0)}(x_t)}\right)\right\} \nn\\
V_t^{(1,2)}(x_t)&=&(1-\rho^2)\frac{\mu^4 c^2}{4\gamma k^3}
\left\{
\frac{(1-Y_{tT})[3m(1-Y_{tT})+2Y_{tT}~x_t]}{2m^2 (X_T^{(0)}(x_t))^2}+
\frac{1}{m^3}\ln\left(\frac{Y_{tT}~ x_t}{X_T^{(0)}(x_t)}\right)
\right\} \nn\\
V_t^{(1,3)}(x_t)&=&\frac{3\rho \mu^3 c^3}{2\gamma k^3}\left\{
\frac{(1-Y_{tT})}{2m^2(X_T^{(0)}(x_t))^2}\bigl[m(1-Y_{tT})-2Y_{tT}~x_t\bigr]-\frac{2}{m^3}\ln
\left(\frac{Y_{tT}~x_t}{X_T^{(0)}(x_t)}\right)\right.\nn\\
&&\hspace{15mm}\left.-\frac{(1-Y_{tT})}{2m^2(X_T^{(0)}(x_t))^3}\Bigl[
5m^2(1-Y_{tT})^2+9m(1-Y_{tT})(Y_{tT}x_t)+2(Y_{tT}x_t)^2\Bigr]\right\}\nn
\eea
\bea
V_t^{(2,2)}(x_t)&=&-\frac{\rho^2 \mu^4 c^2}{\gamma k^3}\left\{\frac{(1-Y_{tT})[3m(1-Y_{tT})+2Y_{tT}x_t]}{2m^2 (X_T^{(0)}(x_t))^2}
+\frac{1}{m^3}\ln\left(\frac{Y_{tT}~x_t}{X_T^{(0)}(x_t)}\right)\right\}\nn\\
V_t^{(2,3)}(x_t)&=&\rho(1-\rho^2)\frac{9\mu^5c^3}{4\gamma k^4}
\left\{\frac{(1-Y_{tT})\bigl[11m^2(1-Y_{tT})^2+15m(1-Y_{tT})(Y_{tT}x_t)
+6(Y_{tT}x_t)^2\bigr]}{6m^3 (X_T^{(0)}(x_t))^3}\right.\nn\\
&&\hspace{30mm}\left.+\frac{1}{m^4}\ln\left(\frac{Y_{tT}~x_t}{X_T^{(0)}(x_t)}\right)\right\}~\nn 
\eea
\bea
V_t^{(3,3)}(x_t)&=&-\frac{3\rho^3 \mu^5 c^3}{\gamma k^4}\left\{
\frac{(1-Y_{tT})\Bigl[11m^2(1-Y_{tT})^2+15m(1-Y_{tT})(Y_{tT}x_t)
+6(Y_{tT}x_t)^2\Bigr]}{6m^3 (X_T^{(0)}(x_t))^3}\right.\nn\\
&&\hspace{30mm}\left.+\frac{1}{m^4}\ln\left(\frac{Y_{tT}~x_t}{X_T^{(0)}(x_t)}\right)\right\}~\nn
\eea
and 
\bea
Z_t^{(0,1)}(x_t)&=&-\frac{\mu^2 c}{2\gamma k}\frac{1-Y_{tT}}{\sqrt{x_t}(X_T^{(0)}(x_t))}\nn \\
Z_t^{(0,3)}(x_t)&=&-\frac{3\mu^2 c^3}{2\gamma k^2}\frac{(1-Y_{tT})^2}{\sqrt{x_t}
(X_T^{(0)}(x_t))^3}\Bigl[m(1-Y_{tT})+2Y_{tT}~x_t\Bigr] \nn \\
Z_t^{(1,2)}(x_t)&=&-\frac{\rho \mu^3 c^2}{\gamma k^2}\frac{(1-Y_{tT})^2}{\sqrt{x_t}(X_T^{(0)}(x_t))^2}\nn \\
Z_t^{(1,3)}(x_t)&=&(1-\rho^2)\frac{3\mu^4 c^3}{4\gamma k^3}\frac{(1-Y_{tT})^3}{\sqrt{x_t}(X_T^{(0)}(x_t))^3} \nn \\
Z_t^{(1,4)}(x_t)&=&-\frac{6 \rho \mu^3 c^4}{\gamma k^3}\frac{(1-Y_{tT})^3[2m(1-Y_{tT})+5 Y_{tT}~x_t]}{\sqrt{x_t}(X_T^{(0)}(x_t))^4}~\nn 
\eea
\bea
Z_t^{(2,3)}(x_t)&=&-\frac{3\rho^2 \mu^4 c^3}{\gamma k^3}
\frac{(1-Y_{tT})^3}{\sqrt{x_t}(X_T^{(0)}(x_t))^3}~\nn \\
Z_t^{(2,4)}(x_t)&=&\rho (1-\rho^2)\frac{9 \mu^5 c^4}{\gamma k^4}\frac{(1-Y_{tT})^4}{\sqrt{x_t}(X_T^{(0)}(x_t))^4}~\nn \\
Z_t^{(3,4)}&=&-\frac{12 \rho^3 \mu^5 c^4}{\gamma k^4}\frac{(1-Y_{tT})^4}{\sqrt{x_t}(X_T^{(0)}(x_t))^4}~.
\eea
It then specifies the optimal strategy $\pi_t^*$ in (\ref{optport}) up to the fourth order of vol-of-vol.
\end{proposition}

\section{Numerical Comparison to the Exact Solution}
In \cite{Zaripho}, it is shown that the Cole-Hopf transformation 
allows the closed form solution for our problem.
We define $K_t=e^{\eta V_t}$ with some constant $\eta\in \mathbb{R}$.
Then, the dynamics of $K$ is given by
\bea
dK_t/K_t&=&\left(\frac{\gamma \eta}{2}(1-\rho^2)+\frac{\eta^2}{2}\right)Z_t^2 dt\nn\\
&&+\left\{\frac{\mu\eta}{\sqrt{X_t}}\rho Z_t-\frac{\mu^2\eta}{2\gamma}\frac{1}{X_t}\right\}dt+\eta Z_tdB_t~.
\eea
Thus, by choosing $\eta^*=-\gamma(1-\rho^2)$ one can eliminate the quadratic term. By 
defining $Q_t=\eta^*K_tZ_t$, the above equation becomes
\bea
dK_t=\left(\frac{\mu\rho}{\sqrt{X_t}}Q_t-\frac{\mu^2\eta^*}{2\gamma}\frac{K_t}{X_t}\right)dt+Q_t dB_t~,
\eea
which is a linear FBSDE with terminal value $K_T=1$.

Now, let us introduce a new measure $\mathbb{P}^*$ for which Brownian motion is 
related to that in the original measure $\mathbb{P}$ by
\bea
dB_t^*=dB_t+\frac{\mu\rho}{\sqrt{X_t}}dt~.
\eea
Then, we have
\bea
dK_t=\frac{\mu^2 (1-\rho^2)}{2X_t}K_tdt+Q_tdB_t^*~,
\eea
which can be integrated easily.
Thus, the solution of the original FBSDE is given by
\bea
V_t=-\frac{1}{\gamma(1-\rho^2)}\ln\left\{
\mathbb{E}^{\mathbb{P}^*}\left[\left.\exp\left(
-\frac{\mu^2}{2}(1-\rho^2)\int_t^T \frac{ds}{X_s}\right)\right|\calf_t\right]\right\}
\label{eq-exact}
\eea
where $X$ follows
\bea
dX_t=k(n-X_t)dt+c\sqrt{X_t}dB_t^*
\eea
under the new measure, where the adjusted mean $n$ denotes $n=m-\rho \mu c/k$.

The diffusion part $Z$ is given by
\be
Z_t=c\sqrt{X_t}\left(\frac{\part V_t}{\part x_t}\right)
\ee
where the partial derivative by the initial value can be easily estimated by 
taking the delta of $V$ relative to the shift of $x_t$. Although $Z$ can also be written with 
a Malliavin derivative of $X$, the higher order terms $\propto 1/X_s^2~(s>t)$ 
and the dynamics of stochastic flow makes it difficult to achieve stable results
of Monte Carlo simulation when it is directly applied to its expression.

{\it Remark: Note that the Cole-Hope transformation cannot always be used to derive exact solutions 
in more generic situations, such as
cases including multi-dimensional risk factors, time or state dependent correlation parameters, e.t.c..
Our scheme can be extended easily, at least in principle,  for these cases, too. }

\subsection{Numerical Comparison}
We now numerically estimate the the solution in Eq.(\ref{eq-exact}) by Monte Carlo (MC) simulation.
In order to guarantee the positivity of $X$, we use the implicit Milstein scheme~\cite{MC}:
\bea
X(t_n)=\frac{X(t_{n-1})+k n \Delta t+c\sqrt{X(t_{n-1})}\xi_n \sqrt{\Delta t}+\frac{1}{4}c^2\Delta t \bigl(\xi_n^2-1\bigr)}
{1+ k \Delta t}
\eea
where $(t_n)_{n\geq 1}$ is equally spaced time grids and $\Delta t=t_{n}-t_{n-1}$. $(\xi_n)_{n\geq 1}$
is a sequence of independent random variable with standard normal distribution $\bold{N}(0,1)$.  
We have run $1$-million plus $1$-million antipathetic scenarios with 
step size $\Delta t=0.005$ to obtain the numerical estimate of $V_0$ in Eq.~(\ref{eq-exact}).
We have compared it to the results of our asymptotic expansion up to the third order of vol-of-vol.
Furthermore, for the diffusion part, we have run another $(1+1)$-million scenarios 
to obtain $V_0$ with the initial value of $X$ shifted by a small amount  $\Delta x_0=5\times 10^{-4}$
to estimate $(\part V_0/\part x_0)$. We have  then multiplied it by $c\sqrt{x_0}$ to obtain the numerical 
estimate of $Z_0$. We have compared it with the analytical approximation up to the fourth order of vol-of-vol.

Table~\ref{eg1} gives the comparison of $V_0$ with $m=6.25\%$ and $c=5\%$, which corresponds to
roughly $\sqrt{m}=25\%$ implied volatility of the risky asset with 
$c/\sqrt{m}=20\%$ vol-of-vol in log-normal terms. 
The each column represents the maturity $T$, the result of MC simulation, its standard deviation,
$\ep$-0th, $\ep$-1st, $\ep$-2nd and $\ep$-3rd order approximation, respectively.  
All the parameters used are provided in the caption.
One can see that the approximation is quite accurate even for 10-year maturity.
Table~\ref{Zeg1} gives the comparison of $Z_0$ with the same parameters in Table~\ref{eg1}.
The column with the label "err" gives the expected error of $Z_0$ 
implied from the standard deviation in the estimation of $V_0$.
Consistently with the convergence of $V_0$, one can see that the diffusion part 
converges nicely to the estimated true value of $Z_0$.

Since the analytical approximation is given by the power series of vol-of-vol "$c$", one can 
expect that its performance deteriorates when the larger $c$ is used. One can see this
in Table~\ref{eg6} where we have used $m=6.25\%$ and $c=12\%$, which 
corresponds to $\sqrt{m}=25\%$ and $c/\sqrt{m}=48\%$. Especially 
for longer maturities, one can observe that the zero-th and first order expansions significantly over/under estimate 
$V_0$. Although $\ep$-2nd and 3rd order
approximations still provide reasonable estimation of the true value in this example,
one needs higher order expansions or some new devise to improve the approximations for larger values of $c$, in general.
For example, it would be better to introduce the expansion parameter $\del$ also in the drift term of $X$ 
to avoid the appearance of small parameters in denominators of the resultant formulas. 
These possibilities may be pursed in a separate paper~\footnote{After submitting this work,
we have developed the new Monte Carlo scheme inspired by the branching diffusion method
to bypass the needs of the asymptotic expansion by allowing simulation of underlying state processes 
directly~\cite{FT2012}.}.
Table~\ref{Zeg6} gives the corresponding comparison for $Z_0$ with the same parameters used in Table~\ref{eg6}.

\begin{table}[!htb]
\small
\hspace{5mm}
\begin{tabular}{|c|c|c|c|c|c|c|}
\hline 
maturity (yr) & $Z$-MC (\%) & err (\%)& $\ep$-0th (\%) &$\ep$-1st (\%) &$\ep$-2nd (\%) &$\ep$-3rd (\%) \\ \hline
1 & -4.293   & 0.015  & -4.442  &  -4.250  & -4.258    & -4.258    \\ \hline
2 & -7.860   & 0.042  & -8.470  &  -7.725  & -7.785    & -7.783   \\ \hline
3 & -10.760  & 0.065  & -12.055 &  -10.471 & -10.661   & -10.650   \\ \hline
4 & -13.094  & 0.082  & -15.205 &  -12.594 & -12.998   & -12.972   \\ \hline
5 & -14.974  & 0.090  & -17.950 &  -14.208 & -14.907   & -14.859   \\ \hline
6 & -16.498  & 0.092  & -20.329 &  -15.420 & -16.480   & -16.403   \\ \hline
7 & -17.740  & 0.096  & -22.384 &  -16.320 & -17.787   & -17.676   \\ \hline
8 & -18.752  & 0.118  & -24.154 &  -16.984 & -18.884   & -18.736   \\ \hline
9 & -19.588  & 0.158  & -25.675 &  -17.469 & -19.812   & -19.625   \\ \hline
10& -20.267  & 0.211  & -26.982 &  -17.820 & -20.603   & -20.377   \\ \hline
\end{tabular}
\caption{A comparison to the MC simulation and asymptotic expansion of $Z$ with parameters:
$m=6.25\%, k=15\%$, $c=5\%, x_0=m, \mu=17\%, \rho=-30\%, \gamma=1$.}
\label{Zeg1}
\end{table}

\begin{table}[!htb]
\small
\hspace{5mm}
\begin{tabular}{|c|c|c|c|c|c|c|}
\hline 
maturity (yr) & $Z$-MC (\%) & std err (\%)& $\ep$-0th (\%) &$\ep$-1st (\%) &$\ep$-2nd (\%) &$\ep$-3rd (\%) \\ \hline
1 & -11.197  & 0.241  & -11.968  & -10.460  & -10.593   & -10.586  \\ \hline
2 & -19.537  & 0.571  & -24.096  & -17.689  & -18.749   & -18.672  \\ \hline
3 & -25.954  & 0.730  & -35.113  & -21.253  & -24.521   & -24.252  \\ \hline
4 & -29.546  & 0.786  & -44.602  & -22.002  & -28.764   & -28.165  \\ \hline
5 & -32.782  & 0.797  & -52.544  & -20.930  & -32.174   & -31.134  \\ \hline
6 & -34.411  & 0.870  & -59.084  & -18.839  & -35.154   & -33.602  \\ \hline
7 & -36.211  & 1.116  & -64.420  & -16.287  & -37.886   & -35.788  \\ \hline
8 & -37.186  & 1.507  & -68.754  & -13.629  & -40.427   & -37.785  \\ \hline
9 & -37.565  & 1.931  & -72.265  & -11.071  & -42.778   & -39.617  \\ \hline
10& -38.079  & 2.382  & -75.108  & -8.723   & -44.925   & -41.285  \\ \hline
\end{tabular}
\caption{A comparison to the MC simulation and asymptotic expansion of $Z$ with parameters:
$m=6.25\%, k=20\%$, $c=12\%, x_0=m, \mu=17\%, \rho=-30\%, \gamma=1$.}
\label{Zeg6}
\end{table}

\begin{figure}[!htb]
	\center{\includegraphics[width=120mm]{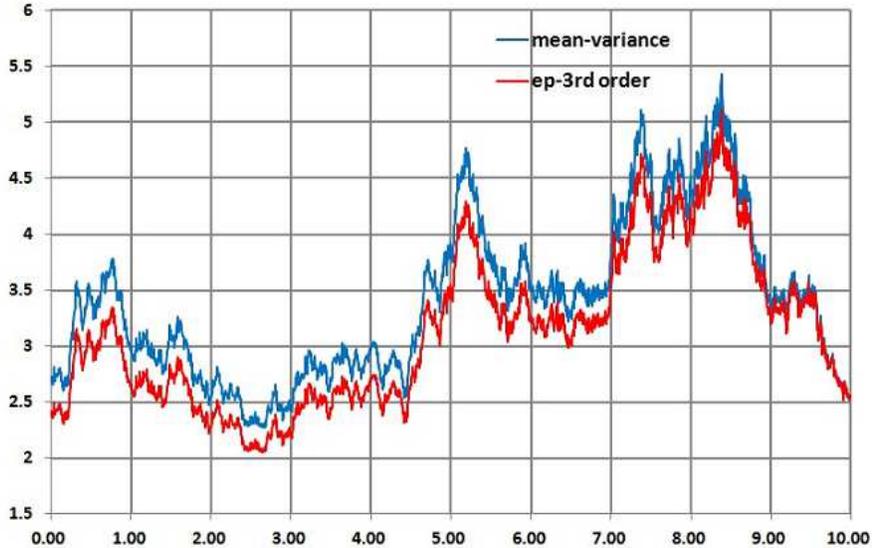}}
	\vspace{-3mm}
	\caption{A sample path each for the mean-variance portfolio and approximated ($\ep$-2nd order) optimal portfolio weight. The used parameters are $m=6.25\%$, $k=15\%$, $c=5\%$, $x_0=m$, $\mu=17\%$, $\rho=-40\%$ and $\gamma=1$.}
	\label{optweights}
\end{figure} 

Lastly, in Figure~\ref{optweights}, we give a sample path each for the 
mean-variance and the approximated $\ep$-3rd order optimal portfolio weight $\pi^*$ with parameters
$m=6.25\%$, $k=15\%$, $c=5\%$, $x_0=m$, $\mu=17\%$, $\rho=-35\%$ and $\gamma=1$ for a $10$-year
investment. One can see that the optimal amount of investment is smaller than that of 
the mean-variance strategy due to the hedging demand. This relationship 
flips the sign when the positive correlation $\rho$ is used.
The difference between the mean-variance and optimal strategies becomes gradually 
smaller as the time comes closer to the maturity as expected.

\section{Conclusion}
In this work, we have studied the optimal portfolio problem 
in an incomplete market with stochastic volatility that is not perfectly hedgeable. 
 We have applied the newly developed 
perturbative methodology combined with standard asymptotic 
expansion technique and derived the explicit solution of 
the corresponding quadratic growth FBSDE up to the 
third order of vol-of-vol for its level and to the fourth order for 
its diffusion component.
The comparison to the exact solution shows quite 
encouraging results about its accuracy
even for quite long maturities, such as 10 years.
As long as we know, the existing numerical techniques, such as
regression based Monte Carlo simulations,  seem mostly limited
to short maturities, say, several months to one year.
Furthermore, the great advantage of our method is its ability to provide
explicit expressions of the optimal portfolios or hedging strategies,
which obviously have great importance for the practical use.

In contrast to the Cole-Hopf transformation, our method can be applied 
to much more generic setups with multi-dimensional risk factors,
which, we expect, will open real possibilities to obtain explicit 
expressions of optimal portfolios and hedging strategies in incomplete and/or 
constrained markets with realistic assumptions. 
This will be addressed in separate works in the future.

\appendix
\section{Numerical results for the "level" component $V$}

\begin{table}[!htb]
\small
\hspace{5mm}
\begin{tabular}{|c|c|c|c|c|c|c|}
\hline 
maturity (yr) & $V$-MC (\%) & std err (\%)& $\ep$-0th (\%) &$\ep$-1st (\%) &$\ep$-2nd (\%) &$\ep$-3rd (\%) \\ \hline
1 &  23.061 & 0.0003 & 23.539  & 23.035 & 23.049 & 23.049 \\ \hline
2 &  45.844 & 0.0008 & 47.769  & 45.671 & 45.787 & 45.783 \\ \hline
3 &  68.197 & 0.0013 & 72.510  & 67.691 & 68.086 & 68.067 \\ \hline
4 &  90.067 & 0.0016 & 97.630   & 88.997 & 89.919 & 89.868 \\ \hline
5 & 111.455 & 0.0018 & 123.031 & 109.560& 111.313 & 111.207 \\ \hline
6 & 132.397 & 0.0018 & 148.639 & 129.398& 132.317 & 132.128 \\ \hline
7 & 152.938 & 0.0019 & 174.401 & 148.552& 152.987 & 152.685 \\ \hline
8 & 173.128 & 0.0023 & 200.278 & 167.076& 173.377 & 172.932 \\ \hline
9 & 193.011 & 0.0031 & 226.239 & 185.028& 193.537 & 192.918 \\ \hline
10& 212.630 & 0.0041 & 252.263 & 202.468 & 213.508 & 212.686 \\ \hline
\end{tabular}
\caption{A comparison to the MC simulation and asymptotic expansion of $V$ with parameters:
$m=6.25\%, k=15\%$, $c=5\%, x_0=m, \mu=17\%, \rho=-30\%, \gamma=1$.}
\label{eg1}
\end{table}

\begin{table}[!htb]
\small
\hspace{5mm}
\begin{tabular}{|c|c|c|c|c|c|c|}
\hline 
maturity (yr) & $V$-MC (\%) & std err (\%)& $\ep$-0th (\%) &$\ep$-1st (\%) &$\ep$-2nd (\%) &$\ep$-3rd (\%) \\ \hline
1 & 24.340  & 0.0020 & 25.461  & 23.896  & 23.992  & 23.988 \\ \hline
2 & 49.550  & 0.0048 & 54.541  & 47.232  & 48.090  & 48.035 \\ \hline
3 & 73.840  & 0.0061 & 86.046  & 68.269  & 71.261  & 71.038 \\ \hline
4 & 96.840  & 0.0066 & 119.177 & 86.407  & 93.441  & 92.870 \\ \hline
5 & 118.640 & 0.0066 & 153.398 & 101.609 & 114.899 & 113.761 \\ \hline
6 & 139.490 & 0.0072 & 188.350 & 114.097 & 135.960 & 134.016 \\ \hline
7 & 159.560 & 0.0093 & 223.792 & 124.189 & 156.901 & 153.913 \\ \hline
8 & 179.030 & 0.0125 & 259.561 & 132.223 & 177.919 & 173.660 \\ \hline
9 & 198.030 & 0.0161 & 295.551 & 138.520 & 199.144 & 193.404 \\ \hline
10& 216.650 & 0.0200 & 331.688 & 143.364 & 220.643 & 213.235 \\ \hline
\end{tabular}
\caption{A comparison to the MC simulation and asymptotic expansion of $V$ with parameters:
$m=6.25\%, k=20\%$, $c=12\%, x_0=m, \mu=17\%, \rho=-30\%, \gamma=1$.}
\label{eg6}
\end{table}

\section{Formulas for $X$'s Asymptotic Expansion}
\label{X_expansion}
We assume $(u>t)$ throughout this section. The value $x_t$ is 
defined as the initial condition at time $t$ by
\be
x_t=X_t^{(\del)}~.
\ee

\subsection{$\del$ 0th order}
The relevant equation becomes deterministic in this case:
\bea
dX_u^{(0)}=k(m-X_u^{(0)})du
\eea
and thus 
\bea
X_u^{(0)}=Y_{tu}x_t+m(1-Y_{tu})
\eea
where we have defined
\be
Y_{tu}=\exp\Bigl(-k(u-t)\Bigr)~.
\ee

\subsection{$\del$ 1st order}
Since we have
\bea
d(\part_\del X_u^{(\del)})=-k(\part_\del X_u^{(\del)})du+
\left(c\sqrt{X_u^{(\del)}}+\frac{1}{2}\del c(X_u^{(\del)})^{-\frac{1}{2}}(\part_\del X_u^{(\del)})\right)dB_u
\eea
which yields
\bea
dD_{tu}=-kD_{tu}du+c\sqrt{X_u^{(0)}}dB_u
\eea
and hence
\bea
D_{tu}=c\int_t^u Y_{us}\sqrt{X_s^{(0)}}dB_s~.
\eea
\subsection{$\del$ 2nd order}
Since we have
\bea
&&d(\part_\del^2 X_u^{(\del)})=-k(\part_\del^2 X_u^{(\del)})du\nn \\
&&\quad +\left\{c(X_u^{(\del)})^{-\frac{1}{2}}(\part_\del X_u^{(\del)})-
\frac{1}{4}\del c(X_u^{(\del)})^{-\frac{3}{2}}(\part_\del X_u^{(\del)})^2+\frac{1}{2}
\del c (X_u^{(\del)})^{-\frac{1}{2}}(\part_\del^2 X_u^{(\del)})\right\}dB_u\nn
\eea
which yields
\bea
dE_{tu}=-kE_{tu}du+c(X_u^{(0)})^{-\frac{1}{2}}D_{tu}dB_u
\eea
and hence
\bea
E_{tu}=c\int_t^u Y_{us}(X_s^{(0)})^{-\frac{1}{2}}D_{ts}dB_s~.
\eea
\subsection{$\del$ 3rd order}
We have
\bea
&&d(\part_\del^3 X_u^{(\del)})=-k(\part_\del^3 X_u^{(\del)})du \nn \\
&&\quad+\Bigl\{ -\frac{3}{4}c(X_u^{(\del)})^{-\frac{3}{2}}(\part_\del X_u^{(\del)})^2+
\frac{3}{2}c(X_u^{(\del)})^{-\frac{1}{2}}(\part_{\del}^2 X_u^{(\del)}) \nn \\
&&\quad+\frac{3}{8}\del c (X_u^{(\del)})^{-\frac{5}{2}}(\part_\del X_u^{(\del)})^3
-\frac{3}{4}\del c (X_u^{(\del)})^{-\frac{3}{2}}(\part_\del X_u^{(\del)})(\part_\del^2 X_u^{(\del)})\nn \\
&&\quad +\frac{1}{2}\del c (X_u^{(\del)})^{-\frac{1}{2}}(\part_\del^3 X_u^{(\del)})\Bigr\}dB_u
\eea
thus,
\bea
dF_{tu}&=&-kF_{tu}du+\frac{3}{2}c\left\{(X_u^{(0)})^{-\frac{1}{2}}E_{tu}-\frac{1}{2}(X_u^{(0)})^{-\frac{3}{2}}D_{tu}^2
\right\}dB_u
\eea
and then
\bea
F_{tu}=\frac{3}{2}c\int_t^u Y_{us}\Bigl\{(X_s^{(0)})^{-\frac{1}{2}}E_{ts}-\frac{1}{2}(X_s^{(0)})^{-\frac{3}{2}}D_{ts}^2\Bigr\}dB_s~.
\eea

\subsection{Relevant expectation values}
It is easy to check that
\bea
\mathbb{E}\Bigl[D_{tu}\Bigr|\calf_t\Bigr]=
\mathbb{E}\Bigl[E_{tu}\Bigr|\calf_t\Bigr]=
\mathbb{E}\Bigl[F_{tu}\Bigr|\calf_t\Bigr]=0~.
\eea
On the other hand, we have
\bea
d(D_{tu})^2&=&2D_{tu}dD_{tu}+d\langle D_{t\cdot}\rangle_u\nn \\
&=&-2kD^2_{tu}du+c^2X_u^{(0)}du+2cD_{tu}\sqrt{X_u^{(0)}}dB_u
\eea
and hence
\bea
D^2_{tu}(x_t)&:=&\mathbb{E}\Bigl[D^2_{tu}\Bigr|\calf_t\Bigr]=
c^2\int_t^u e^{-2k(u-s)}X_s^{(0)}(x_t)ds\nn \\
&=&\frac{c^2}{2k}(1-Y_{tu})\Bigl[(1-Y_{tu})m+2Y_{tu}x_t\Bigr]~.
\eea
By following the similar procedures, it is easy to confirm that
\bea
\mathbb{E}\Bigl[D^3_{tu}\Bigr|\calf_t\Bigr]=\mathbb{E}\Bigl[D_{tu}E_{tu}\Bigr|\calf_t\Bigr]=0~.
\eea


\begin{thebibliography}{99}
\bibitem{Bismut}
Bismut, J.M. (1973). "Conjugate Convex Functions in Optimal Stochastic Control," J. Political
Econ., 3, 637-654.

\bibitem{Carmona}
Carmona (editor) (2009). "Indifference Pricing," Princeton University Press.

\bibitem{crepey}
Cr\'epey, S. (2011). "A BSDE Approach to Counterparty Risk under Funding Constraints,"
Working paper, Universit\'e d'Evry.

\bibitem{Duffie}
Duffie, D., Huang, M. (1996). 
"Swap Rates and Credit Quality," Journal of Finance,
Vol. 51, No. 3, 921.

\bibitem{ElKaroui}
El Karoui, N., Peng, S.G., and Quenez, M.C. (1997). "Backward stochastic 
differential equations in finance," Math. Finance $\bold{7}$ 1-71.


\bibitem{asymmetric_collateral}
Fujii, M., Takahashi, A. (2010). "Derivative pricing under Asymmetric and 
Imperfect Collateralization and CVA,"
CARF Working paper series F-240, available at
http://ssrn.com/abstract=1731763.


\bibitem{analytic_FBSDE}
Fujii, M., and Takahashi, A. (2011). "Analytical Approximation for 
non-linear FBSDEs with Perturbation Scheme," forthcoming 
in International Journal of Theoretical and Applied Finance.

\bibitem{FT2012}
Fujii, M., and Takahashi, A. (2012). "Perturbative Expansion Technique for Non-linear FBSDEs
with Interacting Particle Method," CARF working paper series F-278,
available at http://ssrn.com/abstract=2038740.

\bibitem{Imkeller1}
Hu, Y., Imkeller, P., and M\"uller, M. (2005). "Utility Maximization 
in Incomplete Markets," The Annals of Applied Probability, Vol. 15, No. 3,
1691-1712.

\bibitem{Imkeller2}
Horst, U., Hu, Y., Imkeller, P., R\'eveillac A., and Zhang, J.
(2011). "Forward-backward Systems for Expected Utility Maximization,"
available at arXive:1110.2713.

\bibitem{MC}
Kahl, C., Jackel, P. (2006). "Fast Strong Approximation Monte Carlo schemes 
for Stochastic Volatility Model," Quantitative Finance, Vol.6~(6), 513-536.


\bibitem{KT}
Kunitomo, N. and Takahashi, A. (2003).
"On Validity of the Asymptotic Expansion Approach
 in Contingent Claim Analysis,"
Annals of Applied Probability, 13, No.3, 914-952.

\bibitem{Ma}
Ma, J., and Yong, J. (2000). "Forward-Backward Stochastic Differential Equations and their
Applications," Springer.

\bibitem{P-Peng}
Pardoux, E., and Peng, S. (1990). "Adapted Solution of a Backward Stochastic Differential Equation,"
Systems Control Lett., 14, 55-61.

\bibitem{Skiadas_review}
Skiadas, C., 2008, "Dynamic Portfolio Choice and Risk Aversion,"
Chapter 19 of Handbooks in OR \& MS, Vol. 15.



\bibitem{STT}
Shiraya, K., Takahashi, A., and Toda, M. (2009).
"Pricing Barrier and Average Options under Stochastic Volatility Environment," 
CARF Working Paper F-242, available at http://www.carf.e.u-tokyo.ac.jp/workingpaper/,
forthcoming in Journal of Computational Finance.

\bibitem{T}
Takahashi, A. (1999).
"An Asymptotic Expansion Approach to Pricing Contingent Claims," 
Asia-Pacific Financial Markets, 6, 115-151.

\bibitem{asymptotic3}
Takahashi, A., Takehara, K., and Toda, M. (2011). "A General Computation Scheme for a High-Order Asymptotic Expansion Method,"
CARF Working Paper F-242, available at http://www.carf.e.u-tokyo.ac.jp/workingpaper/.

\bibitem{TY}
Takahashi, A. and Yoshida, N. (2004). 
``An Asymptotic Expansion Scheme for Optimal Investment Problems,''
Statistical Inference fo

\bibitem{Zaripho}
Zariphopoulou, T. (2001). "A Solution Approach to Valuation 
with Unhedgeable Risks," Finance and Stochastics.
\end{thebibliography}
\end{document}